\documentclass[a4paper,12pt]{article}
\usepackage[cp1251]{inputenc}
\usepackage[russian]{babel}

\newcommand{\be}{\begin{equation}}
\newcommand{\ee}{\end{equation}}
\begin{document}

{\large\bf\noindent Dark energy domination in the local flow of
giant galaxies}

\vspace{0.3cm}

\noindent{A.D.~Chernin$^{1}$,  N.V.~Emelyanov$^1$, I.D.
~Karachentsev$^2$}

\vspace{0.2cm} \noindent $^1$Sternberg Astronomical Institute,
Moscow University, Moscow, 119991, Russia;

\noindent $^2$Special Astrophysical Observatory RAS, Nizhnij
Arhyz, 369167, Russia

\vspace{1cm} A dozen most luminous galaxies at distances up to 10
Mpc from the Local Group are moving away from the group forming
the local expansion flow of giants. We use recent Hubble Space
Telescope data on the local giants and their numerous fainter
companions to study the dynamical structure and evolutionary
trends of the flow. A N-body computer model that reproduces the
observed kinematic of the flow is constructed under the assumption
that the flow is embedded in the universal dark energy background.
In the model, the motions of the flow members are controlled by
their mutual attraction force and the repulsion force produced by
the dark energy. It is found that the dark energy repulsion
dominates the force field of the flow. Because of this, the flow
expands with acceleration quantified with the local deceleration
parameter. The dark energy domination is enhanced by the
environment effect of the low mean matter density on the spatial
scale of 50 Mpc in the local universe. The accelerating expansion
cools the flow and introduces to it a nearly linear
velocity-distance relation with the time-rate that is close to
Hubble's factor of the global cosmological expansion.


\vspace{0.4cm}

\section{Introduction}

Expansion flows of galaxies were discovered and studied on spatial
scales from $\simeq 1$ to $\simeq 300$ Mpc (Sandage 1986; Sandage
et al. 1972, 1999, 2006). The kinematic of the flows follows
rather closely the linear velocity-distance relation $V \simeq H
R$ known as Hubble's law. They are considered cold in the sense
that the deviations from this relation are relatively small and
the corresponding radial velocity dispersion $\sigma$ is low in
the observed flows.

The nearest and best studied (Sandage et al. 1972; Karachentsev et
al. 2002, 2003a, 2009; Tully et al. 2013; Courtois et al. 2013)
example of the cold expansion flows is the local flow of dwarf
galaxies (LFD) around the Local Group at the distances of $\simeq
1-3$ Mpc from the group barycenter. Early observations found the
velocity dispersion to be $\sigma \simeq 60$ km/s in the flow
(Sandage et al. 1972; Sandage 1986; Peebles 1988) which is
significantly lower than theory expectations (Peebles 1980). The
data by Ekholm et al. (1999, 2001) and especially recent high
accuracy observations with the Hubble Space Telescope
(Karachentsev et al. 2002, 2003a, 2009) give even lower value
$\sigma \simeq 30$ km/s. The other key physical parameter of the
flows is the expansion time-rate $H$; it was found (Sandage et al.
2006; see also references therein) to be equal within 15-20\%
accuracy to the global cosmological time-rate (Hubble's factor)
$H_0 \simeq 70$ km/s/Mpc. In the LFD, the mean expansion time-rate
is 72 km/s/Mpc with 10\% accuracy (Karachentsev et al. 2009).

In search for the physical nature of the coldness of the local
flows and their nearly cosmological time-rates, two main acting
factors have been proposed: the universal dark energy background
and the particular local environments of the flows. The studies of
the local dark energy effects (Chernin et al. 2000; Baryshev et
al. 2001; Chernin 2001) assume that the dark energy is represented
by Einstein's cosmological constant $\Lambda$, as in the currently
standard $\Lambda$CDM cosmology. The constant is positive which
implies that the dark energy produces a repulsive "force of
antigravity". The cosmic repulsion acts not only on the global
cosmological distances where it was originally discovered in
observations, but actually everywhere in space. In the volume of
the Local Group ($R \le 1$ Mpc), the dark energy antigravity is
weaker than the gravity produced by the matter mass of the group;
but the antigravity repulsion from the group dominates at the
distances $\simeq 1-3$ Mpc outside group's barycenter in the LFD
area. The antigravity domination makes the flow expand with
acceleration, and it is showed that because of this, the LFD is
cooling with time and its time-rate tends to the cosmological
value $H_0$. These results (Chernin et al. 2000,2004; Baryshev et
al. 2001; Karachentsev et al. 2003b; Teerikorpi et al. 2005) are
consistent with general astronomical considerations by Sandage et
al. (2006) and cosmological simulations by Macci\'o et al. (2005),
Peirani \& de Freitas (2008), Nurmi et al. (2010). The same
physics reveals itself also in several other well studied
expansion flows on the spatial scales of 1-10 Mpc (Chernin et al.
2007a,b,c, 2010, Chernin 2013).

On the other hand, Hoffman et al. (2008), Martinez-Vaquero et al.
(2009) find the local effect of dark energy to be marginal in
constrained N-body simulations of the Local Group. Van de Weigaert
\& Hoffman (2000) argue that the observed properties of the flow
somehow reflect the position of the group in the local universe.
While these simulations reproduce to some degree the local cold
flow they are not able (Aragon-Calvo et al. 2011) to identify its
origin apart from being an intrinsic property of the group
environment. The studies of the environment factor take into
account the void-wall-filament nature of the large-scale cosmic
structure (Klypin et al. 2003; Aragon-Calvo et al. 2011; Einasto
et al. 2010, 2012) and suggest that the acceleration and coldness
of the local flow may be due to the particular geometry and
dynamics of the expanding low-density cosmological wall in which
the Local Group resides.

In the present paper, we study the motions of the most luminous
nearest galaxies seen at the distances up to 10 Mpc from the
barycenter of the Local Group. The galaxies and their numerous
fainter companions have recently been observed (Karachentsev et
al. 2013, 2014) with the Hubble Space Telescope (HST). Two of the
local giants -- the Milky Way and the Andromeda Nebula -- are
moving towards each other inside the Local Group, while a dozen
others are moving away from the group with the radial velocities
from 100 to 900 km/s. This is the local flow of giants (LFG), as
we refer to it hereafter. Contrary to the LFD, there is no single
dominating matter concentration in the LFG and each of the flow
members contributes more or less equally to system's total matter
mass. We describe the flow motions with the "$\Lambda$N-body"
computer model which is a N-body model accounting as well for the
universal dark energy background in which the flow is embedded. A
combination of the HST data and the model enables us to recognize
the dynamical structure of the LFG and the flow major evolutionary
trends. Both local effect of dark energy and environment factor
acting in the LFG are in the focus of the discussion.

In Sec.2, basic data on the LFG are given in brief; the
$\Lambda$N-body model is presented in Sec.3; the dynamics and
evolution of the flow are analyzed in Secs.4,5, and the results
are summarized in Sec.6.

\section{LFG: Basic data}

The observation data on the local giant galaxies and their
companions are presented in the recently published Updated Nearby
Galaxy Catalogue (UNGC) by Karachentsev et al. (2013) (see also
Karachentsev 2005; Karachentsev et al. 2014 and Karachentsev \&
Kudrya 2014). The catalog contains systematic and homogeneous data
on coordinates, distances, radial velocities and other basic
physical parameters of about 800 galaxies at the distances up to
10 Mpc. More than 300 UNGC galaxies at these distances have been
observed with the HST for more than 300 HST orbits. The unique
resolution available in the HST observations allowed to use the
Tip of the Red Giant Branch (TRGB) method for precise measurements
of distances to more then 300 nearby galaxies with the accuracy of
10\%. Precise data on the radial velocities of the galaxies have
also been compiled in the UNGC. Modern accurate data on distances
and other parameters of nearby galaxies are also accumulated in
the Extragalactic Distance Database (Tully et al. 2008) and in the
Database of the Local Volume Galaxies (Kaisina et al. 2012).

According to Karachentsev et al. (2013,2014), there are 15 nearby
giant galaxies at the distances up to 10 Mpc: the Milky Way and
the Andromeda Nebula (gravitationally bound in the Local Group),
galaxies M81, NGC5128, IC342, NGC253, NGC4736, NGC5236, NGC6946,
M101, NGC4258, NGC4594, NGC3115, NGC3627, NGC3368 -- see Table 1
(MNRAS 449, 2069, 2015).

Each of the giants of Table 1 is actually the main galaxy of a
group (similar to the Local Group) which includes the galaxy
itself and its extended dark matter halo together with companion
galaxies therein. The mass $M$ in the table is the total orbital
mass of the group. The UNGC data are used to estimate the masses
of the groups via motions of their 351 less massive companions
(Karachentsev \& Kudrya 2014). The distance $R$ of a giant and its
radial velocity $V$ in the table are calculated relative to the
barycenter of the Local Group. The Supergalactic coordinates of
the giants are also given in Table 1.

The nearest to the Local Group is the M81 galaxy with the distance
of 3.6 Mpc. Its recession velocity (relative to the Local Group
barycenter) is 100 km/s, the lowest one in the LFG. The mass of
the galaxy is $5\times 10^{12}M_{\odot}$ (we give here somewhat
rounded numbers; the exact figures with the error bars may be
found in the paper by Karachentsev \& Kudrya, 2014).

The most distant is the galaxy NGC 3368 at 10.4 Mpc with the
radial velocity 740 km/s which is the second largest velocity in
the LFG. Its mass is $2\times 10^{13} M_{\odot}$ which is the
second largest mass in the flow. The most massive one is probably
the galaxy NGC 4594 at 9.3 Mpc; its mass is $3\times 10^{13}
M_{\odot}$, but with the error of $2\times 10^{13} M_{\odot}$, the
largest error in the data. The typical mass estimation error for
the LFG is not more than 30-40\%. The NGC 4594  is the second most
distant galaxy with the highest radial velocity of 890 km/s.

As is seen from Table 1, the spatial distribution of 15 LFG giants
reveals a concentration to the Supergalactic plane (see also
McCall 2014); the region is referred to as the Local Pan-Cake
(Karachentsev et al. 2013). According to Peebles \& Nusser (2010)
and McCall (2014), 10 giants at distances $\simeq 1-8$ Mpc form
the flattened Local Sheet. The Local Pan-Cake/Sheet looks like
part of the large-scale Cosmic Web of walls/sheets/pan-cakes,
galaxy clusters, filaments and voids. The Virgo and Fornax
clusters at the distances less than 20 Mpc are the closest to the
LFG most massive elements of the web. Their masses are $8 \times
10^{14} M_{\odot}$ and $1 \times 10^{14} M_{\odot}$, respectively;
the distances are 17 and 20 Mpc, the radial velocities are 975 and
1410 km/s (Karachentsev et al. 2010).

Using the UNGC and the data by Karachentsev \& Kudrya (2014), we
may make rough quantitative estimates of the gross characteristics
of the LFG as a whole system. The mean distance in the sample is
$<R> \simeq 7$ Mpc, the mean radial velocity is $<V> \simeq 400$
km/s, and the mean velocity-to-distance ratio is $ <H>  \simeq 65$
km/s/Mpc which is equal -- within 10\% accuracy -- to the
cosmological value $H_0 \simeq 70$ km/s/Mpc. The rms radial
velocity dispersion is $\sigma \simeq 100$ km/s; the value is
considerably less than the mean velocity $<V>$ and the
characteristic "virial" velocity of the system $V_{vir} = [\frac{G
M_{tot}}{<R>} ]^{1/2} \simeq 300$ km/s. Here $M_{tot} = 8\times
10^{13} M_{\odot}$ is the sum of the matter (dark matter and
baryons) masses of the flow member. The mean matter density in the
sphere of 10 Mpc in radius $<\rho_M> \simeq 1 \times 10^{-30}$ g
cm$^{-3}$, which is about 0.3 the mean cosmological matter density
$\rho_{0} \simeq 3 \times 10^{-30}$ g cm$^{-3}$. It indicates
(Karachentsev et al. 2012, 2014, Karachentsev \& Kudrya 2014) that
the volume where the LFG resides is an underdensity region in the
local universe. The whole size of the underdensity region reaches
the distances of about 50 Mpc from the Local Group (Vennik 1984;
Tully 1987; Magtesyan 1988; Bahcall et al. 2000; Crook et al.
2007; Makarov \& Karachentsev 2011).

\section{$\Lambda$N-body model}

With the physical characteristics presented in Sec.2, the LFG
looks similar to the LFD or any other local expansion flows: it is
rather cold and its mean expansion time-rate is near the global
time-rate $H_0$. However there are significant differences between
the LFG and the LFD. Indeed, the galaxies of the LFD are moving in
the external gravity-antigravity field produced by the Local Group
and the dark energy, while the mutual gravity of the bodies is
practically negligible (Sec.1). The dynamics of the LFG is also
controlled by the gravity-antigravity force field, but in contrast
to the LFD, there is no single dominating matter concentration in
the LFG; each of its member galaxies contributes more or less
equally to the total matter mass of the system, and the gravity is
due to the mutual attraction of the bodies. The model of the flow
we suggest is the $\Lambda$N-body model which treats the LFG as a
N-body system embedded in the dark energy background.

\subsection{Local antigravity force}

The local dynamical effects of the universal dark energy are
actively studied (see for review Chernin 2001, 2008, 2013; Byrd et
al. 2008, 2012) since the discovery of the dark energy in
cosmological observations. Here we give some basic relations
needed for the description of the force field in our model. The
relations come from General Relativity with non-zero Einstein's
cosmological constant $\Lambda$ and the currently standard
$\Lambda$CDM cosmology. The dark energy is considered as a
continuous medium with the density $\rho_{\Lambda} =
c^2\Lambda/(8\pi G) > 0$ that is positive and constant in space
and time in any reference frame (here $G$ is the Newtonian
gravitational constant and $c$ is the speed of light). The
currently adopted value of the dark energy density $\rho_{\Lambda}
\simeq 0.7 \times 10^{-29}$ g/cm$^{3}$. The dark energy produces
the cosmic repulsion, or antigravity, which dominates the dynamics
of the observed universe as a whole.

The dark energy as a medium is characterized by the equation of
state (Gliner 1965):
\be p_{\Lambda} = - \rho_{\Lambda} c^2.\ee

\noindent Here $p_{\Lambda}$ is the dark energy pressure. The dark
energy is a vacuum-like medium: it cannot serve as a reference
frame, and it is co-moving to any matter motion -- similarly to
trivial emptiness (Gliner 1965).

In General Relativity, the effective gravitational density is
determined by both density and pressure of any fluid:
\be \rho_{eff} = \rho + 3 p/c^2.\ee

\noindent The effective density of dark energy, $\rho_{\Lambda
eff} = \rho_{\Lambda} + 3 p_{\Lambda}/c^2 = - 2 \rho_{\Lambda} <
0$, is negative, and it is because of this that dark energy
produces the repulsion, or antigravity, not gravity.

General Relativity indicates also that the passive gravitational
density is the sum $\rho_{pass} = \rho + p/c^2$ for any fluid. The
value is zero for the dark energy. According to the equivalence
principle, the passive gravitational mass is equal to the inertial
density. Thus, the inertial density of the dark energy is zero.
This implies that the dark energy is affected neither by the
external gravity of matter nor by its own antigravity.

We assume here that this interpretation of the dark energy may be
borrowed from $\Lambda$CDM cosmology and applied to relatively
small, non-cosmological spatial scales of $\simeq 1-300$ Mpc, no
matter that the isotropic cosmology itself is not valid on these
local scales. We will describe the local dynamical effects of dark
energy in terms of Newtonian mechanics; it is possible because the
velocities of the local flows of galaxies are small compared to
the speed of light, and the spatial differences in the
gravity-antigravity potential are much smaller (in absolute value)
compared to the speed of light squared.

In this weak field approximation, Einstein's law of universal
antigravitation says: any body in the universe is affected by the
repulsive force which is proportional to the dark energy density
$\rho_{\Lambda}$ and to the distance $R$ of the body from the
origin of the adopted reference frame and acts along the direction
from the origin:
\be F_{\Lambda} = {\frac{8\pi}{3}} G \rho_{\Lambda} R. \ee

\noindent The relation of Eq.3 gives the force per unit mass of
the body, i.e. acceleration, in the projection on the body
radius-vector. It may be rewritten in the form $F_{\Lambda} =
H_{\Lambda}^2 R$, where
\be H_{\Lambda} = [{\frac{8\pi}{3}} G \rho_{\Lambda}]^{1/2} = 61
\;\; km/s/Mpc \ee

\noindent is the "universal time-rate" (see below). An exact
analytical solution for two-body problem on the dark energy
background with the antigravity force of Eq.3 has recently been
given by Emelyanov \& Kovalyov (2013). The same Eqs.3,4 are used
below for the N-body computer solutions.

\subsection{Equations of motion}

In our $\Lambda$N-body model, the LFG (together with the two
nearest clusters or without them) is treated as a non-relativistic
isolated conservative system of point-like masses interacting with
each other via Newton's mutual gravity and undergoing Einstein's
antigravity (Eq.3) produced by the dark energy background in which
the giants and the clusters are embedded. The equations of motion
for the system have the following form (see, for comparison, books
by Duboshin 1975, and Roy 1978 on N-body problem without the dark
energy):
 \be  \frac{d^2 x_i}{dt^2}= G{\sum\limits_{j=1}^N}^\prime m_j
\frac{x_j-x_i}{r_{ij}^3}+H_{\Lambda}^2 x_i, \ee
\be \frac{d^2 y_i}{dt^2}=G{\sum\limits_{j=1}^N}^\prime m_j
\frac{y_j-y_i}{r_{ij}^3}+H_{\Lambda}^2 y_i, \ee
\be \frac{d^2 z_i}{dt^2}=G{\sum\limits_{j=1}^N}^\prime m_j
\frac{z_j-z_i}{r_{ij}^3}+H_{\Lambda}^2 z_i, \ee

\noindent where
\be  r_{ij}=\sqrt{(x_j-x_i)^2 + (y_j-y_i)^2 + (z_j-z_i)^2}.\ee

Here $x,y,z$ are the Cartesian coordinates with the origin in the
barycenter of the LFG; $m_i$ is the mass of the body of the number
$i$, the symbol prime $(\prime)$ at the summation symbol means the
absence of the item with $j=i$. In Eqs.5-7, Newton's gravity force
depends on the distance between gravitating bodies, while
Einstein's force depends only on the distance of the body from the
coordinate origin.

The number of the LFG bodies is 13, since the Local Group with the
Milky Way and the Andromeda Nebula is considered as one mass;
another close gravitationally bound binary galaxy, NGC5236 and
NGC5128, is also considered as one mass (see Table 1). Together
with the two clusters, the total number of the bodies $N = 15$, in
the basic model. In the model for the LFD (Sec.5) the total number
of bodies N = 35.

A numerical integration of Eqs.5-7 is performed with the use of
Everhart's (1974) standard computer method with the automatic
choice of the integration step.

\subsection{Initial conditions}

The initial conditions of the motions are specified at the present
moment of cosmic time $t = t_0 = 13.7$ Myr; these are the observed
positions of the bodies and their measured radial velocities
re-calculated to the barycenter of the Local Group.

Note that observations provide us with the radial velocities of
the galaxies, but say nothing about their tangential (transverse)
velocities. This is an obvious flaw in any dynamical model of
extragalactic astronomy where the full velocity vector is needed
for correct formulation of initial conditions. Only for the
Andromeda Nebular an estimate of the transversal velocity has
recently become available (van der Marel \& Guhathakurta 2008).
Being cognizant completely  about the situation, we assume here
for simplicity that the transversal velocities are zero in the
initial conditions for the model (see also Secs.4,5 below).

\subsection{Ten integrals}

As is well-known (see again Duboshin 1975 and Roy 1978), the
N-body equations of motion have 10 first integrals; they
correspond to the spacetime symmetries of the equations which
imply respective conservation laws. The following quantities must
be constant in the N-body motion on the dark energy background:
\be a_x=\sum\limits_{i=1}^N m_i x_i, \;\;\;
 a_y=\sum\limits_{i=1}^N m_i y_i, \;\;\;
 a_z=\sum\limits_{i=1}^N m_i z_i, \;\;\; \ee
\be b_x=\sum\limits_{i=1}^N m_i \dot{x}_i, \;\;\;
 b_y=\sum\limits_{i=1}^N m_i \dot{y}_i, \;\;\;
 b_z=\sum\limits_{i=1}^N m_i \dot{z}_i, \;\;\;\ee
\be c_x=\sum\limits_{i=1}^N m_i (y_i \dot{z}_i - z_i\dot{y}_i) ,
\;\;\;
 c_y=\sum\limits_{i=1}^N m_i (z_i \dot{x}_i - x_i\dot{z}_i) , \;\;\;
 c_z=\sum\limits_{i=1}^N m_i (x_i \dot{y}_i - y_i\dot{x}_i) , \;\;\;
\ee
\be
  E =\frac{1}{2}\sum\limits_{i=1}^N m_i(\dot{x}_i^2 + \dot{y}_i^2 + \dot{z}_i^2 ) -
    \frac{1}{2}{\sum\limits_{i=1}^N \sum\limits_{j=1}^N}^\prime
    \frac {G m_i m_j}{r_{ij}} -
    \frac{1}{2}H_{\Lambda}^2\sum\limits_{i=1}^N m_i({x}_i^2 +
    {y}_i^2 + {z}_i^2).
\ee

The energy conservation law of Eq.12 says that the total
mechanical energy of the system is the sum of the total kinetic
energy of the bodies and their total potential energy in the
gravity-antigravity force field. In the barycentric coordinates
used in in Eqs.5-8, the values of $a_x, a_y, a_z, b_x, b_y, b_z $
must be equal to zero.

The ten integrals of Eqs.9-12 are employed to ensure accuracy in
our computer integrations: the integrals are computed and their
deviations from the initial values are calculated at the end of
each computation session. The integrals prove to be conserved
unchanged, and the deviations are found to be zero with all the
accuracy of the number presentation in the computer. As for
efficiency of integration, the solution of $\Lambda$N-body problem
with, say, N = 15-40 on the cosmic time interval of 15 Gyr takes
less than 5 seconds of computing time.

\section{LFG dynamics and evolution}

We have performed computer integration of Eqs.5-7 in two versions.
One of them reproduces the system which includes all the giant
galaxies of the flow and also two nearest clusters of galaxies
(Virgo and Fornax) outside the flow. This is our basic
$\Lambda$N-body model for the LFG discussed below in this section.
In the other version, the system includes the LFG galaxies only.
Comparing the models, we may clarify and evaluate the effect of
the environment factor in the dynamics of the flow -- see below
Sec.5. Both models cover the time interval of 15 Gyr, from the
cosmic age at $t = 10$ Gyr in its past to the present age of $t =
13.7$ Gyr and further to $t = 25 $ Gyr in the future. The history
of the flow is traced back in time to narrower spatial
configurations of the LFG bodies that still allow their
approximate treatment as point-like masses; but deeper in the
past, the distances between the bodies would be comparable with
their proper sizes. As for the future, the integration is limited
by the condition of non-relativistic velocities of expansion. The
description of the system for 15 Gyr is seemingly long enough to
recognize its dynamical structure and major trends of its
evolution.

\subsection{Phase trajectories}

In the basic $\Lambda$N-body model ($N = 15)$, the state of the
system is given by its position in the 6N-dimension phase space at
any instant in the cosmic time interval from 10 to 25 Gyr. For
visualization of the results, a two-dimensional $V-R$ hypersurface
of system's phase space is used, where $V \equiv \dot R$ is the
radial velocity of a body and $R$ is its radial distance -- see
Fig.1 (in MNRAS 449, 2069, 2015). For this figure, the coordinate
transformation is made from
system's barycentric ($x,y,z$) coordinates of Eqs.5-7 to the
supergalactic ($X,Y,Z$) coordinates used in Table.1. Also the
radial distances $R$ and radial velocities $V$ are re-calculated
to the reference frame of the Local Group barycenter. In Fig.1,
the present (observed) state of the system is showed by the set of
black dots. These data are used as "initial conditions" for the
integration of the equations of motion. The thick lines in Fig.1
are the phase trajectories of 15 bodies during 15 Gyr of the flow
evolution.

It is seen from Fig.1 that the lengths of the phase trajectories
are relatively short for the bodies with low initial velocities
and long for the high-velocity galaxies and the clusters. All the
trajectories demonstrate clearly the major evolutionary trend of
the flow: with the distance growth, they converge to the straight
line $V = H_{\Lambda} R$ going from the coordinate origin. As a
result, the mean radial velocity dispersion decreases with the
distance growth, and the flow gets increasingly regular and cold.
Indeed, the spread of the phase trajectories around the line $V =
H_{\Lambda} R$ is systematically decreases with the increase of
the distances. For instance, the width of the LFG trajectory bunch
decreases by a factor of 3 when the distance $R$ increases from
7.5 to 17 Mpc. This is a clear quantitative measure of the cooling
effect in the expansion flow.

The behavior of the phase trajectories in Fig.1 shows also that
the flow tends asymptotically to the state in which the
trajectories follow the linear velocity-distance relation, $V =
H_{\Lambda} R$. In this asymptotic state, the flow expansion
time-rate $ V/R = H_{\Lambda} = 61$ km/s/Mpc (see Sec.3). This
time-rate is constant and depends on the dark energy density only.
Its value is not far from the cosmological time-rate $H_0 \simeq
70$ km/s/Mpc found in global observations (see also below).

On the other distance limit, $R < 5$ Mpc, relatively small, but
obvious deviations from the straight linearity are seen relatively
in the phase diagram. These deviations and also intersections of
some trajectories there reflect the dynamical effect of the mutual
gravitational interactions of the flow bodies. No signs of body
binary interactions are seen at the large ($R > 10$ Mpc) distances
in Fig.1. This shows that, with the distance growth, the mutual
gravity of the flow bodies becomes weak compared to the repulsion
antigravity of the dark energy, and the gravity effect vanishes
eventually in the limit of large distances. Thus, the
velocity-distance diagram for the LFG indicates that the dark
energy dominates the flow dynamics at these distances.

\subsection{Asymptotic state}

The dynamics of the flow in the limit of large distances where the
mutual gravity of the bodies may be neglected is illustrated by an
exact analytical solution for non-gravitating particles moving
along the radial trajectories in the force field produced by the
dark energy alone. The force of the dark energy antigravity is
given by Eqs.3,4, and in this limit, the equation of motion for
any flow body in the reference frame of the Local Group barycenter
has the form:
\be \ddot R = \frac{8\pi G}{3} \rho_{\Lambda} R = H_{\Lambda}^2 R.
\ee

\noindent The first integral of the equation
\be \dot R^2 = H_{\Lambda}^2 R^2 + 2E, \ee

\noindent where $\dot R = V$ is the radial velocity, $E$ is a
constant which is the total mechanical energy of the particle
which may be zero, negative or positive for various trajectories
of the flow. When $R$ goes to infinity, Eq.14 gives:
\be V  \rightarrow H_{\Lambda} R, \;\;\;  R \rightarrow \infty.
\ee

\noindent It is the same asymptotic evolutionary state as is also
seen in Fig.1 in the large distance limit: the linear
velocity-distance relation with the universal time-rate $V/R =
H_{\Lambda}$. In this limit, the distances of the flow bodies grow
exponentially with time: $R(t) \propto \exp{H_{\Lambda}t}$.

The asymptotic time-rate $H_{\Lambda}$ of Eq.15 is known for the
global cosmological expansion. Indeed, according to the
$\Lambda$CDM model, the dark energy dominates the global dynamics
at the cosmic redshifts $z < z_{\Lambda} \simeq 0.7$ and the
cosmic time $t > t_{\Lambda} \simeq 7$ Gyr. Since this epoch, the
dark energy domination is getting stronger with time, and in the
limit of large times, $t >> t_{\Lambda}$, the
Friedmann-Lema$\hat{i}$tre metric tends eventually to the de
Sitter metric. In this limit, the Friedmann equation for the
cosmological expansion is reduced to the form
\be \ddot a(t) = H_{\Lambda}^2 a (t), \ee

\noindent where $a(t)$ is the cosmological scale factor. This
equation is similar to Eqs.13 above, and its solution is similar
to Eqs.14,15 for the flow asymptotic state: $\dot a/a =
H_{\Lambda}, a(t) \propto \exp{H_{\Lambda}t}$. Thus, the temporal
asymptotic of the global cosmological expansion is exactly the
same as the spatial asymptotic of the local expansion flow in the
$\Lambda$N-body model for the LFG. It is because of this
similarity of the global and local that the observed time-rate of
the LFG is near the observed value of the Hubble factor, and the
both are close to the universal time-rate $H_{\Lambda}$.

\subsection{Spatial trajectories and distances}

Alongside with the phase trajectories, the $\Lambda$N-body model
for the LFG gives the spatial trajectories of the flow giants
bodies in the projection to the Supergalactic ($Z = 0$) plane --
see thick lines in Fig.2 (MNRAS 449, 2069, 2015). As is seen from
the figure, the LFG is expanding monotonically: the distances of
the bodies are increasing during all the time of computation, and
the mutual distances between the bodies are also increasing. Some
of the trajectories are slightly curved at small distances; most
of them look like almost straight lines at relatively large
distances. This indicates again that the mutual gravity of the
bodies is relatively weak in the flow as a whole. In agreement
with the results of Fig.1, the spatial trajectories of Fig.2 show
that the LFG was rather regular even 2-3 Gyr ago and it is
evolving now towards an almost perfectly regular asymptotic state
at which mutual binary interactions of the flow members are
insignificant.

The $\Lambda$N-body model reproduces the distances of the flow
bodies from the Local Group barycenter as functions of time from
$t = 10$ Gyr to $t = 25$ Gyr -- see Fig.3 (MNRAS 449, 2069, 2015).
It is seen from the figure that the initially slowest body
increases its distance in 1.7 times (from 3.5 to 6 Mpc) during
this time; meanwhile the fastest one increases its distance in 3.7
times (from 6 to 22 Mpc). So the difference in radial distances
increases in 2.7 times (from 6 to 16 Mpc). This indicates that the
geometrical structure of the LFG evolves in a rather complex way
at relatively small times. But in the asymptotic regime when the
initial conditions are forgot completely, all the distances in the
LFG -- both radial distances $R$ and mutual distances between the
galaxies -- increase with time proportionally to the common scale
factor $\exp{H_{\Lambda}t}$ -- like in the cosmological expansion
(see above).

It should be reminded here that the tangential (transverse)
velocities of the LFG bodies are unknown and we put them to be
zero in the initial conditions for the model (Sec.3). Non-zero
transverse velocities, $V_t \ne 0$, might change the trajectories
of the flow in some way; however they could hardly alternate the
main trends of the flow evolution and especially the asymptotic
state of the flow because the tangential velocities vanish with
the growth of the distances, as the conservation of the total
angular momentum of the flow (Eq.11) indicates. For each
individual body, the product  $V_t R$ is roughly constant in time,
and so $V_t/V \propto R^{-2} \rightarrow 0$ when $V \propto R$ at
the limit $R \rightarrow \infty$. The problem of tangential
velocities in the local flows needs, however, more detailed
studies.

\subsection{Acceleration}

Cosmology was sometimes referred to as  a science of the two
numbers: one of them is Hubble's expansion time-rate $H(t) = \dot
a/a$, and the other is the dimensionless deceleration parameter
which is defined as
\be q(t) = - \frac{\ddot a a}{\dot a^2}. \ee

\noindent Here $a (t)$ is (as above) the cosmological scale
factor. The cosmological deceleration parameter of Eq.17 is
positive for redshifts $z > z_{\Lambda}$ (gravity dominates), it
is zero at $z = z_{\Lambda}$ (zero-gravity moment) and negative
for $z< z_{\Lambda}$ (antigravity dominates) tending to
$q(t)_{\infty} = -1$ in the limit of infinite cosmological
expansion. At the present epoch, the cosmological deceleration
parameter $q_0 \simeq -0.6$.

The time-rate $H(R)$ was introduced above to the LFG as a local
counterpart of Hubble's time-rate in cosmology. In a similar way,
we introduce the deceleration parameter $q(R)$ as a function of
the radial distance $R$ for the local expansion flows (Chernin \&
Teerikorpi 2013):
\be q(R) = - \frac{\ddot R R}{\dot R^2}. \ee

\noindent In the $\Lambda$N-body model of the LFG, the parameter
$q(R)$ is calculated for each of the flow trajectories
individually. The result is presented in Fig.4 (MNRAS 449, 2069,
2915) where the present-day values of the parameter are showed by
black dots. As we see, the deceleration parameter $q(R)$ is
negative for each of the flow bodies at all the distances and time
moments under consideration. It indicates directly that the LFG is
accelerating and the dark energy dominates its dynamics since the
cosmic time $t = 10$ Gyr at least. The mean value of the
deceleration parameter at present $<q(R)_0> \simeq -0.9$.

The deceleration-distance dependence proves to be, generally,
non-monotonic at relatively small distances $R < 10$ Mpc in Fig.4.
It is seemingly due to a well-known sensitivity of few-body or
N-body systems to details of initial conditions. Another cause is
a complex gravity-antigravity interplay in the flow dynamics.
However at larger distances, the initial conditions are mainly
forgot, and the mutual gravity vanishes; as a result, all the
trajectories tend to the asymptotic value $q(R)_{\infty} = -1$
when $R$ goes to infinity. This local asymptotic value is the same
as the cosmological value $q(t)_{\infty}$ (see above).

\section{Local flows: Environment factors}

Klypin et al. (2004), Aragon-Calvo et al. (2011), Nurmi et al.
(2013) and McCall (2014) argue (see also references therein) that
the location of the observed expansion flows is particular and
they may be affected by the large-scale Cosmic Web. Here we study
two environment factors related to the location of the flows:
these are the local underdensity in the large-scale matter
distribution and the nearest most massive overdensities in the
Cosmic Web which are the Virgo and Fornax clusters of galaxies. We
use for these studies the two versions of our computer models for
the LFG and also the $\Lambda$N-body models for the LFD.

\subsection{LFG: Underdensity}

As is mentioned in Sec.2, the mean present-day matter density
$<\rho_M>_0$ estimated by Karachentsev \& Kudrya (2014) for the
spherical volume of 10 Mpc in radius around the LFG is about
one-third the mean cosmological matter density: $<\rho_M>_0 \simeq
0.3\rho_{0}$ (see Sec.2). The underdensity area extends to
distances of 50 Mpc and more being an element of the global Cosmic
Web Klypin et al. (2004) and Aragon-Calvo et al. (2011).

In addition to the densities $\rho_0$ and $<\rho_M>_{0}$, one more
characteristic density is also involved in the local flow dynamics
-- this is the constant in space and time effective gravitating
density of the dark energy $\rho_{eff} = -2 \rho_{\Lambda} = -1.4
\times 10^{-29}$ g cm$^{-3}$ (see Sec.3). This third density is
about 14 times (in absolute value) the mean matter density
$<\rho_M>$ in the LFG area. The ratio $|\rho_{eff}|/<\rho_M>_0$
may serve as a direct quantitative measure of the dark energy
ability to effect the flow. The ratio is high due to the local
underdensity at present and it will increase with the LFG
expansion in the future. In this way, the local antigravity effect
of the dark energy and the environment factor of global
underdensity are acting together in same direction enhancing each
other: they both makes the flow expand with acceleration. In the
limit of large distances, $|\rho_{eff}|/<\rho_M>_0  \rightarrow
\infty$ when $R \rightarrow \infty$.

Is global underdensity needed necessarily for the formation of
typical local expansion flows? To discuss this question, we may
address the example of the LFD around the Local Group and use it
for a comparison with the LFG.

\subsection{LFD: No underdensity}

The LFD is the best studied example of local expansion flows. It
demonstrate all the typical features of these system: nearly
linear velocity-distance relation, low velocity dispersion,
time-rate close to Hubble's global cosmological time-rate (see
Sec.1 and references therein). The flow structure, dynamics and
evolution are clearly demonstrated by the N-body model for the LFD
-- see below.

The flow is seen at distances up to 2.7 Mpc from the center of the
Local Group. The mean matter density in the spherical volume of
2.7 Mpc in radius is near the cosmological matter density, if not
equal to it exactly. Indeed, the mean matter density in this
volume is $M_{LG}/(\frac{4 \pi}{3} R_{LFD}^3) \simeq 0.3 \times
10^{-29}$ g cm$^{-3}$, where $M_{LG} = 3\times 10^{12} M_{\odot}$
is the matter mass of the group, $R_{LFD} = 2.7$ Mpc. This
coincidence is hardly accidental; rather it suggests that the
group and the flow around it occupy the Einstein-Straus vacuole
(see Teerikorpi et al. 2008, Saarinen \& Teerikorpi 2014 and
references therein). Anyway there is no underdensity in the LFD
area.

The characteristic ratio $|\rho_{eff}|/<\rho_M>_0 \simeq
|\rho_{eff}|/\rho_0 \simeq 5$ for the LFD; it is less than that
for the LFG, but the dark energy dominates obviously in the LFD
area. In the case of the LFD, there is no additional environmental
enhancement for the dark energy domination, and nevertheless the
flow is expanding with acceleration as is found by Chernin et al.
(2000, 2004); see also discussion below and Fig.6 (MNRAS 449,
2069, 2015) where the deceleration parameter $q(R)$ for the LFD is
presented which is negative for all the flow bodies. The example
of the LFD suggests that a flow may be accelerating even if the
mean matter density in its area is not less than the cosmological
matter density; but the criterion $|\rho_{eff}|/<\rho_M>_0 \ge 1$
must necessarily be satisfied.

\subsection{LFG as environment factor}

In our $\Lambda$N-body model for the LFD, the flow is surrounded
by the local giants together with the two clusters of galaxies. In
this way, the LFD environment up to distances of 20 Mpc is
included in the consideration. The model is computed with $N = 20
+ 13 + 2$ for dwarfs, giants and clusters, respectively. The
equation of motion of Sec.3 are used together with the
corresponding initial conditions given by the observational data
(Karachentsev et al. 2009, 2014). The masses of the LFD galaxies
are calculated from the data on their red luminosities assuming
the typical mass/luminosity ratio $m/L = 10$, in the Solar units.
The result is presented by the thick lines in Fig.5 (MNRAS 449,
2069, 2015) where the phase structure of the LFD is given (the LFG
and cluster are not showed). The model for the LFD alone ($N =
20$) is also computed -- see thin lines in Figs.5,6 (in MNRAS 449,
2069, 2015).

Weak, but clearly visible differences may be found between these
two models, especially at relatively small distances and
relatively small velocities of the LFD in the diagram of Fig.5. It
is seen from Fig.6 that the deceleration parameter is rather
sensitive to the environment effect at small distances. However
the gross parameters of the accelerating flow as a whole remain
generally unchanged. Thus, the LFD structure and evolution does
not depend significantly on the presence or absence of the giant
galaxies and big clusters around the flow; the dominant role
belongs to the dark energy antigravity in the flow.

\subsection{LFG: Effect of clusters}

In our basic $\Lambda$N-body model for the LFG  with N = 13 + 2
(see Sec.4 above), the two clusters of galaxies -- the Virgo and
Fornax clusters -- are taken into account. In this way, the
gravity of the nearest most massive elements of the large-scale
Cosmic Web is directly included into the dynamics of the flow. To
visualize the effect of the clusters, we performed computations
for a version of the model (N = 13) without the clusters (see the
thin lines in Figs.1-4). The comparison with the basic model (the
thick lines) shows that in the phase space (Fig.1) the clusters do
not affect the trajectories of the bodies with relatively short
initial distances. The motions of only two bodies of the LFG with
the largest (near 10 Mpc) initial distances are obviously affected
as the thin lines indicate this in Fig.1. These two flow
trajectories are shifted to relatively higher velocities from the
asymptotic trajectory $V = H_{\Lambda}R$ which enhances somewhat
the velocity dispersion in the flow. the two flow trajectories are
shifted to relatively higher velocities from the asymptotic
trajectory $V = H_{\Lambda}R$ which enhances somewhat the velocity
dispersion in the flow.

In the picture of the spatial trajectories of Fig.2, the effect
produced by the clusters is also rather weak, but obvious. Indeed,
in the upper part of the diagram, 9 spatial trajectories of the
LFG are inclined towards the trajectory of the Fornax cluster
compared with their position in the model without clusters. No
effect like that is seen from the Virgo cluster. The environmental
effect is weak in Fig.3 as well, and it is seen only in the
fastest trajectories near the cosmic age of 8 Gyr and more.

The high sensitivity of the deceleration parameter to the initial
conditions (as mentioned in Sec.4) is seen in Fig.4. The
difference between the models is much larger here than in
Figs.1-3. The environmental effect looks especially strong at the
distance $R < 10$ Mpc in the $q-R$ diagram. Interestingly enough,
the gravity of the two clusters do not enforce the spread of the
curves in this area of the diagram; on the contrary, it introduces
rather a specific additional regularity to the flow making the
deceleration parameter closer, generally, to the asymptotic value
$q(R)_{\infty} = -1$. Thus, the general evolution trends of the
flow described in Sec.4 look stable and remain roughly unchanged
in the presence or absence of the two closest most massive
overdensities of the Cosmic Web.

\section{Conclusions}

The local expansion flow of giant galaxies has recently been
discovered and studied with the HST observations at the distances
up to 10 Mpc (Karachentsev et al. 2013, 2014, Karachentsev \&
Kudrya 2014). The flow involves dozen most luminous nearby
galaxies together with about 300 their fainter companions. The
flow occupies a flattened volume near the Supergalactic Plane. The
flow is expanding: the giants are moving away from the barycenter
of the Local Group of galaxies with the radial velocities from 100
to about 1000 km/s. The total matter mass of the flow is $M \simeq
8\times 10^{13} M_{\odot}$.

We have developed a set of computer models which treat the local
flow of giant as a N-body system embedded in the perfectly uniform
time-independent background of the dark energy represented by
Einstein's cosmological constant $\Lambda$. In these
"$\Lambda$N-body" models, the dynamical state of the flow and the
trends of its evolution are controlled by the local interplay
between the mutual gravity of the galaxies and the cosmic
antigravity produced by the dark energy. A combination of the
observational data with the computer models leads to following
conclusions:

1) The local flow of giants (LFG) reveals two major observed
features which are common -- according to Sandage (1986), Sandage
et al. (1972, 1999, 2006) -- for expansion flows of various
spatial scales: the nearly linear velocity-distance relation $V
\simeq H R$ and the nearly cosmological mean value of the
expansion time-rate $<H> \simeq 65$ km/s/Mpc.

2) The third major feature of the LFG is its acceleration which is
quantified by the dimensionless deceleration parameter, $q(R)=
\ddot R/(VR)$ borrowed from cosmology. The parameter proves to be
negative for each of the LFG individual galaxies since the cosmic
time $t = 10$ Gyr at least. Its present mean value for the flow as
a whole is $<q(R)> \simeq -0.9$.

3) The negative deceleration parameter means that the flow is
accelerating, and the flow acceleration indicates in its turn that
the flow dynamics is dominated by the antigravity produced by the
dark energy.

4) A direct dimensionless quantitative measure of the dark energy
domination in the LFG is the ratio of the effective gravitating
density (in absolute value) to the mean matter density in the flow
area $|\rho_{eff}|/<\rho_M>$. The present-day value of the ratio
is $\simeq 14$, which is about three times larger then the similar
ratio found earlier for the local flow of dwarf galaxies around
the Local Group at the distances less than 3 Mpc. In the LFG, the
dynamical effect of the dark energy is enhanced by the environment
factor of the low mean matter density existing on the large scale
of 50 Mpc in the local universe.

5) In the limit of large distances  when $R \rightarrow \infty$,
the flow time-rate tends to the value $H_{\Lambda} =
[{\frac{8\pi}{3}} G \rho_{\Lambda}]^{1/2} = 61$ km/s/Mpc which
depends on the dark energy density only. Alongside with this, the
deceleration parameter of each individual galaxies of the flow
goes to the common value $q(R) = -1$. The density ratio
$|\rho_{eff}|/<\rho_M>$ tends to infinity in this limit. The
values $H_{\Lambda}$,  $q_{\Lambda} = -1$ and the infinite density
ratio $|\rho_{eff}|/\rho_M$ are also known as asymptotic
characteristics of the global cosmological expansion in the limit
when $t \rightarrow \infty$. In terms of General Relativity, both
local and global expansion flows have the same asymptotic
4-dimensional state which is the de Sitter space-time.

6) The environment conditions related to the large-scale Cosmic
Web with its most massive nearby matter concentrations (the Virgo
and Fornax clusters of galaxies at distances about 20 Mpc) are
able to modify slightly details of the LFG internal structure.

\section*{Acknowledgements}

We appreciate helpful discussions with G. Bisnovatyi-Kogan, G.
Byrd, Yu. Efremov, P. Teerikorpi, M. Valtonen and A. Zasov.

\vspace{1.5cm}

\section*{References}

\hspace{0.5cm} Aragon-Calvo M.A., Silk J., Szalay, A.S. (2011)
MNRAS 415L, 16

Bahcall N.A., Cen R., Dave R., Ostriker J.P., Yu Q. (2000)

Baryshev Yu., Chernin A., Teerikorpi P. (2001) A\&A 378, 729

Byrd G.G., Chernin A.D., Valtonen M.J. (2007) Cosmology:
Foundations and Frontiers, URSS Pubs., Moscow

Byrd G.G., Chernin A.D., Teerikorpi P., Valtonen M.J. (2012) Paths
to dark energy. De Gruyter, Berlin/Boston

Chernin A.D. (2001) Physics-Uspekhi, 44, 1099

Chernin A.D. (2008) Physics-Uspekhi, 51, 253

Chernin A.D. (2013) Physics-Uspekhi, 56, 704

Chernin A.D., Teerikorpi P. (2013) A\&AT 28, 177

Chernin A.D., Teerikorpi P. (2014) Ast.Rep. 58, 1

Chernin A.D., Teerikorpi P., Baryshev Yu.V. (2000)
(astro-ph//0012021) = Adv. Space Res. 31, 459, 2003

Chernin A.D., Karachentsev I.D., Valtonen M.J. et al. (2004) A\&A
415, 19

Chernin A.D., Karachentsev I.D., Kashibadze O.G. et al. (2007a),
Astrophys. 50, 405

Chernin A.D., Karachentsev I.D., Makarov D.I. et al. (2007b),
A\&AT 26, 275

Chernin A.D., Karachentsev I.D., Valtonen M.J. et al. (2007c),
A\&A 467, 933

Chernin A.D., Karachentsev I.D., Nasonova O.G. et al. (2010) A\&A
520, A104

Chernin A.D., Teerikorpi P., Valtonen M.J. et al. (2012) A\&A.
539, 4

Courtois H.M., Pomarede D., Tully R.B., et al. (2013) AJ 146, 69

Crook A.C., Huchra J.P., Martimbeau N., Masters K.L., Jargett T.,
Macri L.M. (2007) ApJ 655,790

Duboshin G.N. (1975) Celestial mechanics: Basic problems and
methods. Nauka, Moscow (in Russian)

Ekholm T., Teerikorpi P., Theureau G. et al. (1999) A\&A, 347, 99
(1999)

Ekholm T, Baryshev Yu., Teerikorpi P. et al. (2001) A\&A 368, 17

Emelyanov N.V. \& Kovalyov M.Yu. (2013) MNRAS 429, 3477

Gliner E.B. (1966) Sov.Phys. JETP, {\bf 22}, 378

Hoffman Y., Martinez-Vaguero L.A., Yepes G., Gott$\ddot{o}$ber S.
(2008) MNRAS 386, 390

Kaisina E.I., Makarov D.I., Karachentsev I.D., Kaisin S.S. (2012)
AstBu 67, 115

Karachentsev I.D. (2012) Astrophys. Bull. 67, 123

Karachentsev I.D. (2005) AJ 129, 178

Karachentsev I.D., Kudrya Yu.N. (2014) AJ 148, 50

Karachentsev I.D., Sharina M.E., Dolphin A.E., et al. (2002) A\&A,
385, 21

Karachentsev I.D., Makarov D.I., Sharina M.E., et al. (2003a)
A\&A, 398, 479

Karachentsev I.D., Chernin A.D., Teerikorpi P. (2003b),
Astrofizika 46, 491

Karachentsev I.D., Kashibadze O.G., Makarov D.I., Tully R.B.
(2009) MNRAS 393, 1265

Karachentsev I.D., Nasonova O.G. (2010) NMRAS 405, 1075

Karachentsev I.D.,  Makarov D.I. Kaisina E.I. (2013) AJ 145, 101

Karachentsev I.D., Kaisina E.I., Makarov D.I. (2014) AJ 147, 13

Klypin A., Hoffman Y., Kravtsov A.V. et al. (2003) ApJ 596, 19

Macci\`{o} A.V., Governato, F. Horellou, C. (2005) MNRAS 359, 941

McCall M.L. (2014) MNRAS 440, 405

Magtesyan A.P. Astrophys. 28, 150

Makarov D.I. \& Karachentsev I.D. (2011) MNRAS 412, 2498

Martinez-Vaguero L.A., Yepes G., Hoffman Y. et al. (2009) MNRAS
397, 2070

Nurmi P., Hein$\ddot{a}$m$\ddot{a}$ki P., Teerikorpi P., Chernin
A.D. (2010) in "The hidden side of galaxy formation" ed. V.P.
Debattista and C.C. Popescu, AIP Conf. Proc. Volume 1240, 419

Nurmi P., Hein$\ddot{a}$m$\ddot{a}$ki P., Sepp T., et al. (2013)
MNRAS 436, 380

Peebles P.J.E. (1980) The large-scale structure of the universe.
Princeton Univ. Press

Peebles P.J.E. (1988) ApJ 332, 17

Peebles P.J.E. \& Nusser A. (2010) Nature 465, 565

Peirani S., de Freitas Pacheco J.A. (2008) A\&A 488, 845

Roy A.E. (1978) Orbital motion. Hilger, Bristol

Saarinen J. \& Teerikorpi P. (2014) A\&A 568, A33

Sandage A. (1986) ApJ, 307, 1

Sandage A. (1999) ApJ 527, 479

Sandage A. et al. (1972) ApJ 172, 253

Sandage A., Tamman G.A., Saha A., et al. (2006) ApJ 653, 843

Teerikorpi P., Chernin A.D., Baryshev Yu.V. (2005) A\&A 440, 791

Teerikorpi P., Chernin A.D., Karachentsev I.D., Valtonen M.J.
(2008) A\&A 483, 383

Tully R.B. (1987) ApJ 321, 280

Tully R.B., Shaya E.J., Karachentsev I.D., et al. (2008) ApJ 676,
184

Tully R.B., Courtois H.M., Dolbphin A.E., et al. (2013) AJ 146, 86

van de Weygaert R., Hoffman Y. (2000) Cosmic flows workshop 201,
169

van der Marel R.P. \& Guhathakurta P. (2008) ApJ 678, 264

Vennik J. (1984) Tartu Astrofuusika Observatoorium Teated 73, 1

\section*{Figure captions}

Fig.1. Local flow of giants: Velocity-distance diagram. The thin
lines are for the model without the clusters of galaxies. Solid
line is the asymptotic linear relation $V = H_{\Lambda} R$. Black
dots are for the present-day values.

Fig.2. Local flow of giants: The spatial trajectories of the LFG
galaxies in the projection to the Supergalactic ($Z = 0$) plane.
The thin lines are for the model without the clusters. Black dots
are for the present-day values.

Fig.3. Local flow of giants: The distances of the LFG galaxies
from the Local Group barycenter. The thin lines are for the model
without the clusters.

Fig.4. Local flow of giants: The deceleration parameter as a
function of the radial distance of the LFG galaxies. The thin
lines are for the model without the clusters. Black dots are for
the present-day values.

Fig.5. Local flow of dwarfs: Velocity-distance diagram. The thin
lines are for the model without the LFG and clusters. Solid line
is the asymptotic linear relation $V = H_{\Lambda} R$. Black dots
are for the present-day values.

Fig.6. Local flow of dwarfs: The deceleration parameter as a
function of the radial distance of the LFD galaxies. The thin
lines are for the model without the LFG and the clusters. Black
dots are for the present-day values.

\end{document}